\documentclass[12pt]{iopart}
\usepackage{iopams}

\usepackage[latin1]{inputenc}
\usepackage{graphicx}

\begin{document}

\title{Probing statistical properties of Anderson localization with quantum emitters}

\author{Stephan Smolka$^{1,\dag}$, Henri Thyrrestrup$^1$, Luca Sapienza$^1$, Tau B. Lehmann$^1$, Kristian R. Rix$^1$, Luis S. Froufe-Pérez$^2$, Pedro D. Garc\'{i}a$^1$, and Peter Lodahl$^1$}

\address{$^1$ DTU Fotonik, Department of Photonics Engineering, Technical University of Denmark, Building 345V, 2800 Kongens Lyngby, Denmark\\ $^2$ Departamento de F\'{i}sica de la Materia Condensada, and Instituto Nicol\'as Cabrera, Universidad Aut\'onoma de Madrid, E-28049 Madrid, Spain}

\ead{smolka@phys.ethz.ch, pelo@fotonik.dtu.dk}


\begin{abstract} 
Wave propagation in disordered media can be strongly modified by
multiple scattering and wave interference.\ Ultimately the
so-called Anderson-localized regime is reached when the waves
become strongly confined in space. So far, Anderson localization of light has been probed in transmission experiments by measuring
the intensity of an external light source after propagation
through a disordered medium. However, discriminating between
Anderson localization and losses in these experiments remains a major challenge.\ Here we present an
alternative approach where we use quantum emitters embedded in
disordered photonic crystal waveguides as light sources.\
Anderson-localized modes are efficiently excited and the analysis
of the photoluminescence spectra allows to explore their
statistical properties paving a way for controlling Anderson localization in disordered photonic crystals.
\end{abstract}


\section{Introduction}

Wave propagation through a disordered system is usually
described as a random walk~\cite{revmodphys71p313}
where interference effects can be ignored after performing an ensemble average over all configurations of disorder.\ However, if multiple
scattering is very pronounced this approximation fails and wave
interference may survive the ensemble averaging leading to a very different
regime where the wave becomes localized in space~\cite{physrev109p1492}, as
experimentally demonstrated with light~\cite{nature404p850,physrevlett94p113903}, matter~\cite{nature453p891}, and acoustic waves~\cite{naturephys4p945}.\ In this so-called Anderson-localized
regime the intensity of a light wave for a single realization of disorder exhibits very large spatial fluctuations
whereas the ensemble-averaged intensity decays exponentially from the light source
on a characteristic length scale called the localization length, $\xi$.\
Confirming Anderson localization experimentally remains a major
challenge since any optical loss in the system, such as absorption
or scattering out of the structure, also results in an exponential decay
of the intensity profile with an
average loss length, $\l$. In most experimental
situations both effects are simultaneously present and this problem can be
circumvented by analyzing the fluctuations of the transmitted
light  intensity~\cite{nature404p850,naturephys4p945}.\ A drawback
of this approach, however, is that in transmission
experiments only a fraction of the confined modes are
excited, i.e. the ones that have non-vanishing amplitudes at the sample surface~\cite{physrevb82p165103}, thus imposing a limitation on a detailed statistical analysis of Anderson localization.

In this paper, we present a new approach to efficiently excite
Anderson-localized modes by employing the light emission from quantum
dots (QDs) embedded in disordered photonic crystal
waveguides.\ We record the
spatial and spectral intensity fluctuations of the Anderson-localized modes and by analyzing the quality ($Q$) factor distributions of the modes we extract important information about the localization length and average loss length. Based on this analysis we demonstrate that the localization length can be tuned by controlling the degree of disorder of the sample. In an extension of the theory, we show that losses in disordered photonic crystal waveguides are distributed as well. The use of embedded light sources to characterize Anderson localization is expected to open a new route in the research direction within of multiple scattering with possible applications for random lasers~\cite{science320p643} and to enhance the light-matter interaction~\cite{science327p1352}.

\section{Experimental method}

\begin{figure}[t]
    \begin{center}
\includegraphics[width=1\textwidth]{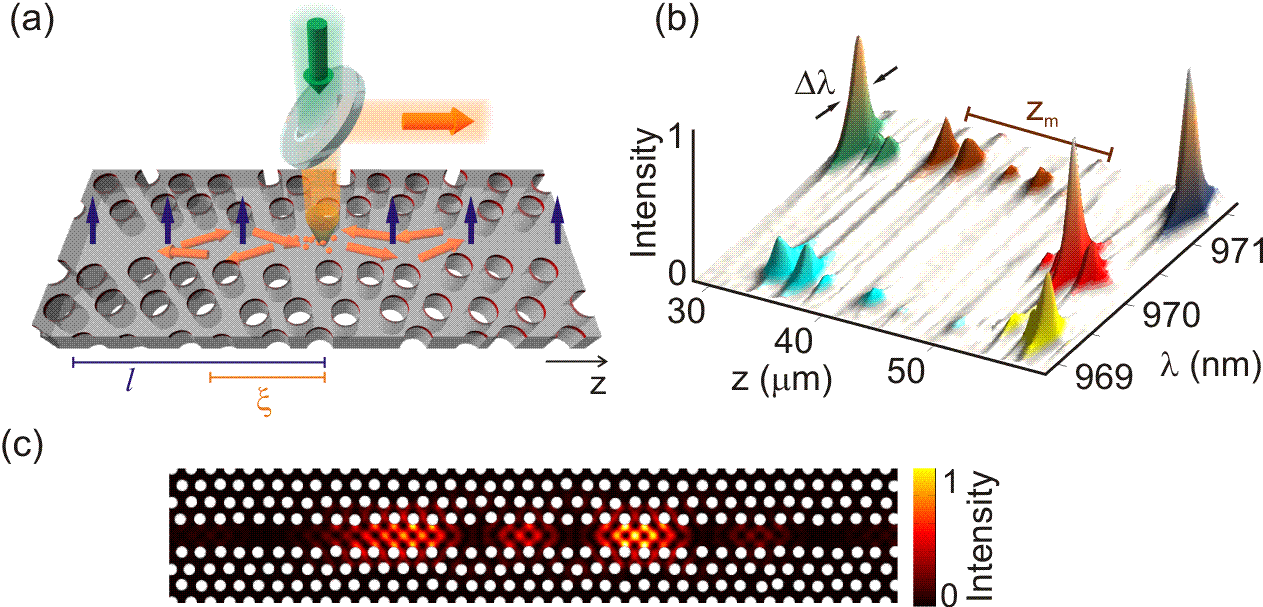} 
\caption{Probing Anderson localization with quantum emitters.\
(a), A continuous wave Ti:Sapphire laser at $\lambda=850\,$nm (sketched
as a green arrow) is focussed onto a disordered photonic crystal
waveguide of length $L=100\,\mu$m  to locally excite randomly
positioned embedded QDs (orange spheres). The emitted QD photoluminescence (sketched as
orange arrows) propagates inside the waveguide and multiple
scattering results in Anderson localization.\ The out-of-plane
scattered light intensity (orange) is collected from the top of
the waveguide using a confocal microscope setup.\ Two loss lengths characterize the transport in the photonic crystal waveguide, i.e. the localization length $\xi$ and the out-of-plane loss length $l$ (not to scale).
(b), Photoluminescence spectra collected while scanning the
excitation and collection objective along a disordered photonic
crystal waveguide with $\delta=3\,\%$ disorder.\ Different modes are illustrated by different color
gradients. The linewidth of a spectral
resonance mode is $\Delta\lambda$, and $z_m$ defines the spatial extension of a mode.
(c), Finite-difference time-domain simulation of the
intensity distribution of an Anderson-localized mode in a
disordered photonic crystal waveguide ($\delta=3\%$) in the high density
of states regime.\ The intensity is plotted on top of the
simulated structure shown in black (GaAs) and white (air). }
  \label{fig01}
      \end{center}
\end{figure}

We investigate photonic crystals consisting of a
triangular periodic lattice of air holes etched in a GaAs membrane
containing a layer of self-assembled InAs QDs in the center with a density
of $\approx 80\,\mu$m$^{-2}$. \ The
periodic modulation of the refractive index opens a frequency band
gap where the in-plane propagation of light is strongly inhibited~\cite{apl93p094102}.\ A W1 waveguide is created by omitting a row of
holes (see Fig.~\ref{fig01}(a)).\ The
height of the membrane is $150\,$nm, the lattice constant is $a=260\,$nm,
the hole radius is $r=78\,$nm, and we investigate various samples all with length
$L=100\,\mu$m.  Light propagation in a photonic crystal waveguide is effectively one-dimensional (1D)
since the 2D band gap efficiently confines light in the
plane of the waveguide, and total internal reflection suppresses out-of-plane
radiation.\ Disorder is induced by randomly varying the hole
positions in the first three rows on each side of the waveguide with a Gaussian distribution
with standard deviations varying between $\delta=0\,\%$ and $6\,\%$ in units of the lattice constant $a$.\ These
perturbations provoke multiple scattering of light in the waveguide resulting in
Anderson localization~\cite{prl99p253901} when $\xi$ becomes smaller
than the sample length, $L$~\cite{Sheng}.\ We note that the underlying structure in the absence of disorder (i.e. an ideal photonic crystal waveguide) is  strongly dispersive giving rise to a modified local density of optical states (DOS), which has been employed for an efficient single-photon source~\cite{prl101p113903}. This modified DOS, however, prevails in the presence of a moderate amount of disorder, which is the situation in the present experiment. This allows studying an interesting interplay between order and disorder where Anderson localization may occur close to the photonic band edge of disordered photonic crystals, as proposed theoretically in a founding work of photonic bandgap materials~\cite{prl58p2486}. As a consequence, a disorder parameter such as the localization length can be controlled to a certain degree by exploiting the underlaying dispersion of the photonic band structure.\

The samples are probed using a confocal
micro-photoluminescence setup for exciting an ensemble of QDs within a diffraction limited region along the waveguide, see Fig.~\ref{fig01}(a).
The emitted light is collected with a microscope objective with a numerical aperture of 0.65.\ The samples are placed in a Helium flow cryostat and cooled down to a temperature of $T=10\,$K. By using a high
excitation power density of $P=2\,$kW/cm$^2$ the QD emission is saturated enabling efficient excitation of Anderson-localized modes over a spectral range of $\lambda=950\pm
50\,$nm. The photoluminescence
is sent to a spectrometer with $50\,$pm resolution and
measured on a CCD camera. The sample position is controlled with
stages providing a spatial resolution of $0.3\,\mu$m. Spatially confined and spectrally separated resonances are clearly visible
in the photoluminescence spectra collected at different positions $z$
along the waveguide (Fig.~\ref{fig01}(b)).\ Neighboring peaks appearing along the waveguide and characterized by the sample central wavelength, $\lambda$,
and linewidth, $\Delta \lambda$, are attributed to the same Anderson-localized mode and plotted in
Fig.~\ref{fig01}(b) with the same color.\ Finite-difference time-domain simulations support the existence of localized modes. Figure~\ref{fig01}(c) displays the calculated intensity at a fixed wavelength showing that light is strongly confined along the waveguide due to the process of multiple scattering. A thorough numerical investigation of Anderson-localized modes in disordered photonic crystal waveguides can be found in Ref.~\cite{prb83p085301}.

\section{Quality factor distribution of Anderson-localized modes}

\begin{figure}[t] 
\centering
\includegraphics[width=1\textwidth]{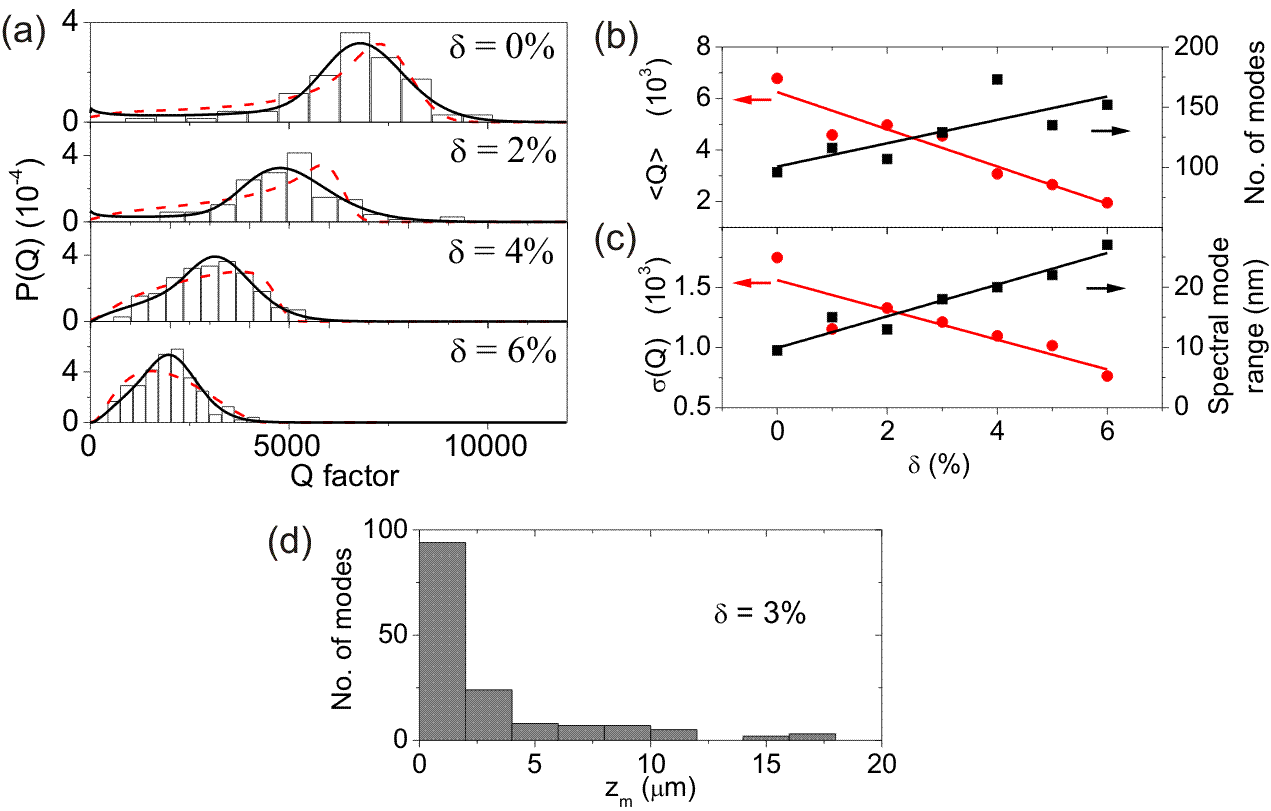} 
\caption{(a), Experimental
$Q$ factor distributions of the observed modes in photonic crystal waveguides (histograms)
for different degrees of induced disorder $\delta$.
The dashed-red curves show the calculated distributions, $p(Q_i^m|\xi,l)$ (see Eq.~(\ref{eq02})) for an average loss length whereas the solid-black curves are with a distribution of loss lengths, $p(Q_i^m|\xi,\mu_l,\sigma_l)$ (see Eq.~(\ref{eq04})).
(b), Average $Q$ factor and number of measured
Anderson-localized modes as a function of $\delta$. (c),
Standard deviation in the experimental $Q$ factor distribution,
$\sigma(Q)$, and spectral range where the Anderson-localized modes
are observed as a function of $\delta$. (d) Maximum distance, $z_m$, between intensity speckles that belong to the same mode, cf. Fig~\ref{fig01}(b). The solid curves in (b) and (c) are guides-to-the-eye.}
\label{fig02}
\end{figure}

Multiple scattering of light is described by a statistical process. As a consequence all characteristic parameters of Anderson-localized modes, i.e. the $Q$ factor or the spatial extent, are distributed and only statistical parameters such as the average or the variance can be predicted. In this section we analyze the distributions of $Q$ factors of the Anderson-localized modes and relate them to the underlying characteristic parameters, i.e., the localization length and the loss length distributions. We observe more than 100 spatially and spectrally
distributed Anderson-localized modes  in each photonic crystal waveguide (Fig.~\ref{fig02}b) at wavelengths in the
region of a high DOS, where the extinction length is the shortest~\cite{physrevb82p165103}.\ The $Q$ factor of the modes, $Q=\lambda/\Delta\lambda$,
is extracted from the intensity spectra collected
along the waveguide by fitting the resonances with lorentzians~\cite{physrevb78p235306}. Since the narrowest $Q$ factors are
influenced by resolution of the spectrometer, all
experimental spectra are first deconvoluted with the measured instrument
response function. The $Q$ factors that are attributed
to the same mode are only
counted once. We observe in Fig.~\ref{fig02}(a) that Anderson localization gives rise to a very broad
distribution of $Q$ factors ranging from $Q=200$ to $Q=10,000$ and notably that the average $Q$ decreases with the amount of introduced disorder. Interestingly the highest $Q$ factors we observe in the Anderson-localized cavities are not far from the values obtained with state-of-the-art engineered nano-cavities with low density of QDs~\cite{nature445p896} despite the fact that our QD density is relatively high. This observation already shows the promising potential of Anderson localization for cavity QED.

Two physical processes contribute to the $Q$ factor of an Anderson-localized mode: light leakage out of the ends of the waveguide due to the finite size of the structure (quality factor $Q_0$) and out-of-plane scattering loss (average quality factor $Q_l$). We have: $Q^{-1}=Q_0^{-1}+Q_l^{-1}$,
where $Q_0$ for 1D multiple scattering is a log-normal distribution determined by the universal parameter $\xi/L$~\cite{jphys38p10761,physreva78p023812}. The out-of-plane light leakage will in general be distributed, i.e. $Q_l$ will depend on the individual configuration of disorder~\cite{prb83p085301}. In the most simple approximation we will assume that only a single average loss length $l$ characterizes the system and have $Q_l=\frac{n\,\pi}{\lambda}\,l$~\cite{Joannopoulos} where $n=3.44$ is the refractive index of GaAs.\ Note that we will go beyond this simple approximation later in the analysis. The light losses are expected to increase with $\delta$ since scattering perturbs the wave vector such that part of the light in the waveguide will have a too small in-plane wave vector component to be trapped by total internal reflection. This mechanism suppresses long scattering paths implying that the $Q$ factor distributions $P(Q)$ are truncated, see Fig.~\ref{fig02}(a).\ The width and mean value of the observed $Q$ factor distributions is determined by the scaled localization length $\xi/L$, and a wide distribution means a small localization length and vice versa. Furthermore, the highest achievable $Q$ is limited by the loss length $l$.

In the following a detailed analysis of the experimental $Q$ factor distributions is presented allowing to estimate $\xi$ and $l$. We can express $P(Q)$ by the log-normal probability distribution of the in-plane $Q$ factors, $P(Q_0)$~\cite{physreva78p023812} that is modified due to the presence of out-of-plane scattering:
\begin{equation}\label{eq00}
P(Q)=\Theta(Q_l-Q)\,\int_0^\infty\,dQ_0\, P(Q_0)\delta\left[Q-(Q_0^{-1}+Q_l^{-1})^{-1}\right],
\end{equation}
where the Heaviside step function, $\Theta$, imposes an upper limit $Q_l$ on the distribution. After evaluating Eq.~(\ref{eq00}) we obtain the conditional likelihood that a 1D disordered medium with a certain localization length $\xi$ and average loss length $l$ supports an Anderson-localized mode with quality factor $Q_i$ %
\begin{equation}
p_1(Q_i|\xi,l)=\exp\left(-\frac{(\mu-\log(\frac{Q_i\,Q_l}{Q_l-Q_i}))^2}{2s^2}\right)
\frac{Q_l\Theta(Q_l-Q_i)}{Q_i\,(Q_i-Q_l)\,\sqrt{2\pi}\,s}\,,\label{eq01}
\end{equation}
where $\mu = (5.9\pm 0.3)\,(\xi/L)^{-0.22 \pm 0.01}$ and $s = (0.4 \pm 0.2)\,(\xi/L)^{-0.59\pm 0.01}.$ Eq.~(\ref{eq01}) is a log-normal distribution characterized by the parameters $s$ and $\mu$ that are related to the localization length via a power law. The explicit expressions for $s$ and $\mu$ were obtained by calculating the in-plane $Q$ factors of Anderson-localized modes in a 1D disordered medium that is composed of a stack of layers with randomly varying real parts of the refractive index. We note that all physical observables in a lossy 1D random medium are determined by the universal parameters $\xi/L$ and $l/\lambda$, i.e. the microscopic details of the medium are indifferent. Therefore, since light propagation in a photonic crystal waveguide is 1D, the stack of randomly varying dieletric layers is an adequate model that is parameterized by the two universal quantities. This model can subsequently be employed to extract the two universal parameters from the experimental data, as will be explained in detail below. We stress that calculations of the actual values of $\xi/L$ and $l/\lambda$ would require full 3D numerical simulations including an appropriate ensemble average over configurations of disorder~\cite{prl103p063903}. We obtain the distribution $P(Q_0)$ by ensemble averaging over eight million different realizations of disorder using an average refractive index of each layer of $n =3.44$ (refractive index of GaAs), a thickness of $L_p=10\,$nm and the sample length of $L=100\,\mu$m. The refractive index is randomly varied by applying a flat distribution within $n\pm\Delta n$, where $\Delta n=(0.22\pm0.03)(\xi/L)^{-0.55\pm 0.01}$. This functional form is obtained after calculating the ensemble-averaged transmission through the stacked layers depending on the sample length, i.e. $\langle\ln T(L)\rangle = -L/\xi$ and for different $\Delta n$. It is evident from Eq.~(\ref{eq01}) that the localization length and the loss length contribute differently to the distribution of Q factors. Experimentally recording the distribution of $Q$ factors therefore enables distinguishing the localization from loss, which is not a priori possible for standard transmission measurements. Since the measured $Q$ factors, $Q_i^m$, have experimental uncertainty, the resultant probability distribution will not be abruptly truncated at $Q_l$ and we account for this uncertainty by convoluting $p_1(Q_i|\xi,l)$ with a normal distribution, $p_2(Q_i^m-Q_i)$, that is centered around $Q_i^m$, i.e.
\begin{equation}\label{eq02}
p(Q_i^m|\xi,l)=\int_0^\infty dQ_i\,p_1(Q_i|\xi,l)\,p_2(Q_i^m-Q_i).
\end{equation}
Assuming that the individual probabilities $p(Q_i^m|\xi,l)$ are independent, the combined probability of measuring a set of $N$ individual $Q$ factors, $\{Q\}=\{Q_1^m,\ldots Q_N^m$\} is $\tilde P(\{Q\}|\xi,l)=\prod_{i=1}^N\,p(Q_i^m|\xi,l)$. In order to estimate the localization length and loss length for a specific set $\{Q\}$ we calculate the inverted conditional probability using the Bayesian theorem~\cite{bayesian}
\begin{equation}\label{eq03}
P(\xi,l|\{Q\})=\frac{\tilde P(\{Q\}|\xi,l)}{P(\{Q\})},
\end{equation}
where $P(\{Q\})$ is a normalization factor. These expressions can be compared to our experimental data of the distribution of Q factors. In Fig.~\ref{fig02}(a) the theoretical $Q$ factor distributions of the form in Eq.~(\ref{eq02}) that give rise to the largest probability $P(\xi,l|\{Q\})$ are plotted and a general good agreement between experiment and theory is observed for all degrees of disorder.

\begin{figure}[t] 
\centering
\includegraphics[width=1\textwidth]{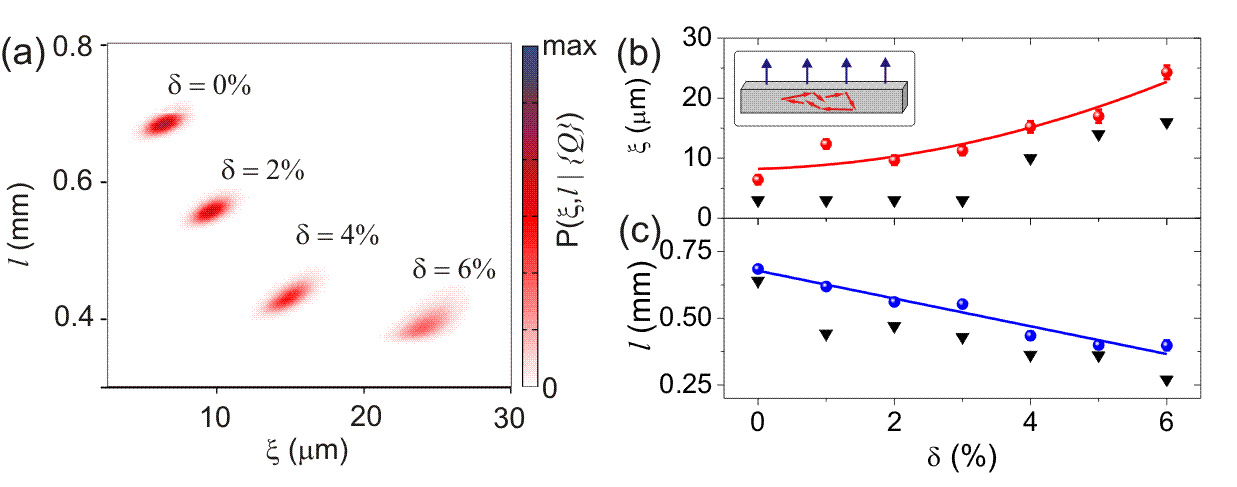}
\caption{(a) Conditional probability $P(\xi,l|\{Q\})$ that a disordered 1D medium with a localization length $\xi$ and average loss length $l$ can be described by the measured $Q$ factor distributions, plotted as a function of $\xi$ and $l$ and for various degrees of disorder. (b), Localization length versus degree of disorder. The red circles are obtained from the data in panel (a) by locating the value where the probability $P(\xi,l|\{Q\})$ is largest.  The black triangles are obtained from $P(\xi,\mu_l,\sigma_l|\{Q\})$ where a distribution of loss lengths were included. Inset: Sketch of the light scattering processes (red arrows) and out-of-plane
scattering (blue arrows). (c), Blue circles (black triangles) are the average loss lengths extracted from $P(\xi,l|\{Q\})$ $(P(\xi,\mu_l,\sigma_l|\{Q\}))$. The solid curves in (b), (c) are guides-to-the-eye.}
\label{fig03}
\end{figure}

Equation~(\ref{eq03}) is a very useful relation since it can be used to extract the localization length and average loss length from the measured $Q$ factor distributions. The dependence of the conditional probability on $\xi$ and $l$ is shown in Fig.~\ref{fig03}(a). We only observe large values of $P(\xi,l|\{Q\})$ in a very restricted range that is strongly dependent on disorder. This enables extracting the localization length and the loss length. The corresponding data are plotted in Fig.~\ref{fig03}(b) that were obtained by averaging over the full spectral range where Anderson-localized modes were observed, cf. Fig. \ref{fig02}(c).\ We extract a localization length that increases with disorder from  $\xi=6\,\mu$m to $\xi=24\,\mu$m, which is shorter than the sample length ($L=100\,\mu$m), thus confirming that the 1D criterion for Anderson localization is fulfilled. We note that the extinction length observed here, which is the total exponential decay length due to contributions from both loss and multiple scattering, is shorter than the extinction lengths measured through standard transmission measurements on samples without light emitters ~\cite{physrevb82p165103}.\ We attribute the difference to the fact that internal light sources can efficiently excite also strongly localized modes that are far away from the waveguide edge, which is much less efficient with an external light source as used in Ref.~\cite{physrevb82p165103}. Quite remarkably, as we vary $\delta$ from $0\%$ to $6\%$, we observe an increase in $\xi$ of about a factor of four, showing that the strongest light confinement takes place in the photonic crystal waveguide when no intentional disorder is added. Unavoidable fabrication imperfections is therefore sufficient to reach the Anderson-localized regime and in fact trap light most efficiently, see Fig.~\ref{fig02}(b).\ The increase of the localization length with disorder is at first sight counterintuitive, since in general increasing the amount of disorder leads to stronger multiple scattering and presumably better localization. However, in a moderately disordered photonic crystal waveguide the DOS is modified near the cutoff of the waveguide mode~\cite{Henri} providing a method to actually control Anderson-localized modes since the localization length is linked to the ensemble-averaged DOS~\cite{physrevb82p165103}. The broadening of the DOS increases with the amount of disorder and consequently the magnitude of the DOS is reduced. This picture is confirmed by the observation that the spectral range of Anderson-localized modes increases with disorder, see Fig.~\ref{fig02}(c). As a consequence also the total number of observed modes increases with disorder, see Fig.~\ref{fig02}(b). Below we will provide a more elaborate analysis where the loss length is distributed. The results from this more complete analysis (also shown in Fig.~\ref{fig03}(b)) indicate that the localization length in fact levels off for small $\delta$ and also predict an even shorter localization length than extracted from the model with a single loss length. The monotonous variation of the localization length with disorder is somewhat unexpected since the combination of order and disorder in photonic crystals could lead to an optimum degree of disorder where localization is most efficient~\cite{prl58p2486,naturephys4p794}. Our results indicate that finding this optimum may require samples with less disorder than the naturally occurring disorder in our current samples. Note that due to the finite statistics in the measurements, the localization length is obtained by averaging over wavelengths, and the above reported scaling with disorder could potentially be highly spectrally dependent. We believe that our data and analysis could inspire thorough numerical investigations of the localization length in disordered photonic crystal waveguides.

The localization length $\xi$ determines the intrinsic ensemble-averaged decay length of the Anderson-localized modes due to confinement by multiple scattering, and is extracted only after accounting for loss processes as presented above. It is instructive also to investigate the directly measured spatial extent of the recorded photoluminescence while scanning along the waveguides, i.e. to extract the length  $z_m$ that is illustrated in Fig.~\ref{fig01}(b). These data are plotted in Fig.~\ref{fig02}(d) for a photonic crystal waveguide with $\delta=3\,\%.$  We observe that $z_m$ varies strongly in the range between  $1\,\mu$m and $18\,\mu$m. These results supplement the localization length measurements and confirm that the modes are indeed strongly localized. We emphasize that the localization length cannot be directly determined from the measurements of $z_m$, since they represent far-field spectra of the photoluminescence obtained by probing at different spatial positions along the waveguide. The average spatial mode extension in general appears to be shorter than the localization length, i.e. we have $\langle z_m\rangle=3\,\mu$m compared to $\xi=10\,\mu$m in the case of $\delta=3\,\%$.  This discrepancy also suggests that the actual localization length is shorter than the values derived from the method described above. Indeed, we assumed that a single loss length was sufficient to describe the out-of-plane scattering, which might be a too simplistic model~\cite{prb83p085301}. Below we will go beyond this assumption by introducing a distribution of loss lengths and indeed in this case a shorter localization length is found.

The dependence of the average loss length on disorder as extracted in Fig.~\ref{fig03}(a) is plotted in
Fig.~\ref{fig03}(c). The loss length is found to decrease with disorder ranging from $l=700\,\mu$m for no intentional disorder to
$l=400\,\mu$m for the largest degree of disorder. Such a behavior has been predicted recently for the role of out-of-plane scattering from the photonic crystal waveguide~\cite{prb83p085301,prl103p063903}.\ The loss length is essential since it limits the highest achievable Q factor leading to a truncation of the distributions of Q factors displayed in Fig~\ref{fig02}(a). Efficient light-matter interaction requires a short localization length and a long loss length, i.e. the samples without engineered disorder are in fact most promising in that respect, as seen in Fig.~\ref{fig03}(b) and (c).

\begin{figure}[t] 
  \centering
  \includegraphics[width=0.7\textwidth]{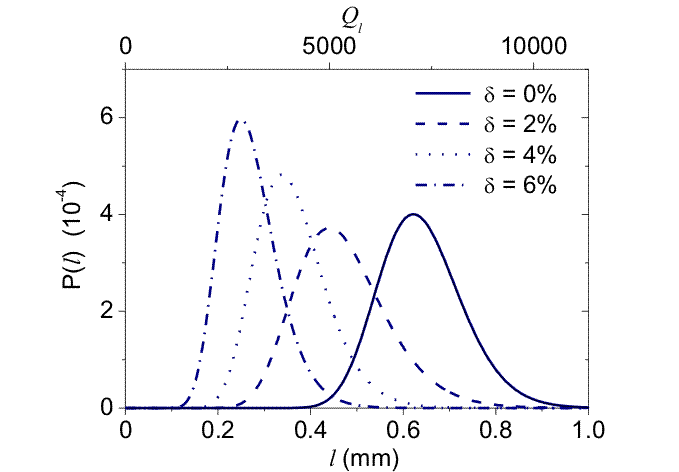} 
\caption{Most likely loss length distributions of the Anderson-localized modes in disordered photonic crystal waveguides for various amounts of disorder.  The top $x$-axis shows the corresponding $Q$ factor.}
  \label{fig05}
\end{figure}

A more elaborate model of a disordered photonic crystal waveguide includes that individual localized modes have different out-of-plane scattering rates leading to a distribution of loss length. This distribution of loss rates has recently been investigated numerically~\cite{prb83p085301}, however no explicit form has to our knowledge been obtained. We will assume this distribution to be log-normal. This choice is motivated by the fact that the light leakage is determined by the coupling between an Anderson-localized mode and all radiation modes. In the absence of radiation modes, the electric field of the localized modes would decay exponentially vertically out of the structure, and we can use a similar argument as for the in-plane $Q$ factors. Assuming that the distribution of vertical decay lengths is Gaussian suggests a log-normal distribution for the overall loss length and consequently for $Q_l$, see Eq.~(\ref{eq01}). The loss distribution is included in our model by integrating over the loss Q-factor distribution in Eq.~(\ref{eq02}), i.e.
\begin{eqnarray}\label{eq04}
\nonumber p(Q_i^m|\xi,\mu_l,s_l)=\\\int_0^\infty dQ_i\int_0^\infty dQ_l\,p_1(Q_i|\xi,l)\,p_2(Q_i^m-Q_i)\,p_3(Q_l|\mu_l,s_l),
\end{eqnarray}
where $\mu_l$ and $s_l$ are the two parameters characterizing the log-normal distribution of the loss Q factor, $p_3(Q_l|\mu_l,s_l)$. The total probability distribution, $P(\xi,\mu_l,s_l|\{Q\})$, can subsequently be calculated in a similar way as explained above leading to  Eq.~(\ref{eq03}). The computational power required to evaluate the multi-dimensional integrals in Eq.~(\ref{eq04}) is very demanding and it is therefore convenient to neglect the uncertainties in the experimentally measured $Q$ factors, i.e. $p_2(Q_i^m-Q_i)=\delta(Q_i^m-Q_i)$. The resultant $Q$ factor distributions that correspond to the largest probability $\max(P(\xi,\mu_l,s_l|\{Q\}))$ are shown in Fig.~\ref{fig02}(a). We observe an even better agreement between experiment and theory as compared to the model where an average loss length was used. The resulting localization lengths are plotted in Fig.~\ref{fig03}(b) and found to increase with disorder. Interestingly the localization length extracted from this improved model is predicted to be shorter (varying from $\xi=3\,\mu$m to $\xi= 16\,\mu$m), which supports our measurements of  $\langle z_m\rangle$ and recent theoretical studies~\cite{prb83p085301}. The loss-length distribution is shown in Fig.~\ref{fig05}. The very wide distributions highlight the importance of extending the model with a distribution of loss lengths. From the distributions we also extract the average loss length $l_d=\exp(\mu_l+s_l^2/2)$ that is displayed in Fig.~\ref{fig03}(c) and found to decrease with disorder similarly to what was observed in the case of a single-loss length as described above. Thus, while a single-parameter loss length is sufficient to predict the correct trends in the localization length and average loss length, it appears essential to include a distribution of loss lengths in a more quantitative analysis. The presence of a distribution of loss lengths is a special property distinguishing disordered photonic crystals from, e.g., non-dispersive waveguide geometries where  a single loss parameter is sufficient~\cite{nature404p850}. Thus, the analysis underlines the complexity of disordered periodic structures where an interplay between order and disorder is responsible for light confinement.

\section{Intensity probability distribution in the Anderson-localized regime}

\begin{figure}[t] 
  \centering
 \includegraphics[width=0.7\textwidth]{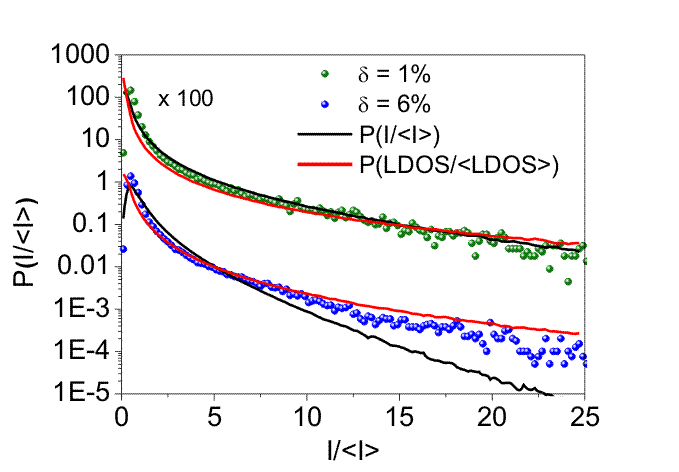} 
\caption{Intensity probability distribution in the
Anderson-localized regime. The measured intensity, $I$, is
normalized to the average intensity, $\langle I\rangle$, and the
data are displayed for $\delta = 1\,\%$ (green dots, scaled by a
factor of 100) and $\delta = 6\,\%$ (blue dots). The black curves
represent the calculated intensity probability distributions, $P(I/\langle I\rangle)$ (see text). The calculated local DOS probability
distributions, $P($LDOS$/\langle$LDOS$\rangle)$, are plotted for the same set of parameters (red
curves).}
  \label{fig04}
\end{figure}

The fluctuations in the emitted light intensity provide important information of the statistical properties of Anderson localization.\  The measured intensity distributions are displayed in Fig.~\ref{fig04} for two different degrees of disorder showing the probability of finding a certain intensity $I$. The probability distributions have very long tails, i.e. very large intensity fluctuations are found for Anderson localization~\cite{nature404p850,physrevlett94p113903,naturephys4p945}.\ To obtain the intensity probability distribution, $P(I/\langle I \rangle)$, we measure the normalized spectral intensity, $I(\lambda,z)/\langle I(z)\rangle$, within the spectral range where the Anderson-localized modes appear. Here $\langle I(z)\rangle$ is the wavelength-averaged intensity at spatial position $z$. For a better signal to noise ratio we scan along the waveguide and extract from these data the total intensity probability distribution. In general, two distinct processes  contribute to the observed intensity fluctuations. First, multiple scattering leads to a random interference pattern and second, multiple scattering causes fluctuations in the local DOS affecting the decay rates of the QDs. We emphasize that the latter contribution to the intensity fluctuations is only present when having light sources embedded in a multiple scattering medium, and have recently been observed experimentally in time-resolved experiments \cite{science327p1352,prl105p013904,optexpress18p6360,Carminati}. The modifications in the local DOS give rise to changes in the intensity that may be pronounced when the QDs are pumped into saturation~\cite{physrevb78p235306}.

The experimental data are compared to calculated intensity and local DOS probability distributions, see Fig.~\ref{fig04} that are obtained by determining the light emission from a point source in a 1D disordered medium using the dyadic Green's function formalism. The Green's function, $G(z_0,z,\lambda)$, describes the electric field at the position $z$ emitted by a monochromatic point source at $z_0$. Since the environment changes on a length scale much smaller than the excitation spot in the experiment a spatial average over one wavelength is carried out~\cite{physrevb65p121101}, $\langle G(z_0,z_0,\lambda)\rangle=\lambda^{-1}\int_{z_0-\lambda/2}^{z_0+\lambda/2} G(z,z,\lambda) dz$. Such a spatial average is needed in describing light emission from an ensemble of emitters, which is the case in the present experiment. The light intensity at the excitation spot for a single realization of disorder is proportional to $|\langle G(z_0,z_0,\lambda)\rangle|^2$ while the local DOS is determined by the imaginary part of the Green's function evaluated at the position of the emitter, i.e. $\Im(\langle G(z_0,z_0,\lambda)\rangle)$. In order to obtain a distribution to simulate the experimental data, we ensemble average over eight million different realizations of disorder applying the same model that we used to determine $P(Q_0)$. The average out-of-plane scattering loss is included in the imaginary part of the refractive indices, i.e. $l=\lambda/(2\,\pi\,\Im(n))$ and the calculations are carried out without any free parameters using the values of the localization length and loss length plotted in Fig.~\ref{fig03}(b), (c) for the single-loss-parameter model.

Figure~\ref{fig04} shows the comparison between the theoretical and the measured intensity distributions. The calculated intensity probability distribution is in good agreement with the experimental data for small intensities while deviations are observed in the tail of the distribution, in particular for large degrees of disorder. Comparing our data to the calculated local DOS distribution we find a surprisingly good agreement. We conclude that likely both the intensity fluctuations and the local DOS fluctuations contribute to the observed intensity speckle pattern. Quantitative agreement between experiment and theory would require a more elaborate theory of the photonic crystal waveguide than the one presented here taking into account the propagation of the light from the photonic crystal waveguide to the detector and the contribution of radiation modes to the local DOS.

\section{Conclusion}
In conclusion, we have presented a novel approach to probe statistical properties of Anderson localization in a disordered photonic crystal waveguide. Using ensembles of QD emitters distributed along the waveguide, Anderson-localized modes are very efficiently excited allowing to study the $Q$ factor distributions and the intensity fluctuations. By analyzing the QD photoluminescence, we recorded a broad distribution of $Q$ factors that is strongly dependent on the induced disorder, and can be explained by changes in the localization length and loss length. Comparing the experimental data with a 1D model for Anderson localization we determined the localization length and found a counterintuitive increase with the amount of disorder, which is attributed to the modified DOS of a photonic crystal waveguide prevailing in the presence of a moderate amount of disorder. We furthermore conclude that the loss of a disordered photonic crystal waveguide is distributed as well, significantly increasing the complexity of the 1D multiple scattering model used to extract universal parameters from the experimental data. The strong dependence of localization and loss length on disorder shows that Anderson localization in disordered photonic crystals is fundamentally different from the case of non-correlated disordered systems. These results possibly open a route to engineering of Anderson-localized modes by controlling the amount and type of disorder, which could significantly improve the performance of quantum electrodynamic experiments in random media \cite{science327p1352}. Finally we recorded the intensity fluctuations of the photoluminescence signal and observed good agreement with our theoretical model using the extracted parameters from the Q factor distribution analysis. The consistency between the two independent set of measurements is an important check of validity of the applied approach proving that a 1D multiple scattering model very successfully describes the behavior of light transport in disordered photonic crystal waveguides.

\section*{Acknowledgements}
We thank P.T. Kristensen for fruitful discussion and gratefully acknowledge financial support from the
Villum Foundation, The Danish
Council for Independent Research (Natural Sciences and
Technology and Production Sciences), and the European Research Council (ERC consolidator grant).

\section*{References} 

\end{document}